\documentclass{article}

\usepackage[margin=1.5in]{geometry}
\usepackage{setspace}
\usepackage{multicol}
\usepackage{graphicx}
\usepackage{amsmath}
\usepackage{amssymb}
\usepackage{lineno}

\begin{document}

\title{An efficient method for sorting and selecting for social behaviour}

\author{
A.~Szorkovszky$^{1,2,*}$, A.~Kotrschal$^{3}$, J.~E.~Herbert Read$^{3,1}$,\\ D.~J.~T.~Sumpter$^{1}$, N.~Kolm$^{3}$ and K.~Pelckmans$^{2}$}

\maketitle

\noindent \hspace*{2.5cm}$^{1}$Mathematics Department, Uppsala University, Sweden\\
\hspace*{2.5cm}$^{2}$IT Department, Uppsala University, Sweden\\
\hspace*{2.5cm}$^{3}$Zoology Department, Stockholm University, Sweden\\
\hspace*{2.5cm}$^*$corresponding author :\\
\hspace*{2.5cm}alexander.szorkovszky@math.uu.se\\

\vspace{0.5cm}





\noindent {\bf Summary}

\begin{enumerate}

\item In social contexts, animal behaviour is often studied in terms of group-level characteristics. One clear example of this is the collective motion of animals in decentralised structures, such as bird flocks and fish schools. A major goal of research is to identify how group-level behaviours are shaped by the traits of individuals within them. Few methods exist to make these connections. Individual assessment is often limited, forcing alternatives such as fitting agent-based models to experimental data.

\item We provide a systematic experimental method for sorting animals according to socially relevant traits, without assaying them or even tagging them individually. Instead, they are repeatedly subjected to behavioural assays in groups, between which the group memberships are rearranged, in order to test the effect of many different combinations of individuals on a group-level property or feature. We analyse this method using a general model for the group feature, and simulate a variety of specific cases to track how individuals are sorted in each case.

\item We find that in the case where the members of a group contribute equally to the group feature, the sorting procedure increases the between-group behavioural variation well above what is expected for groups randomly sampled from a population. For a wide class of group feature models, the individual phenotypes are efficiently sorted across the groups and thus become available for further analysis on how individual properties affect group behaviour. We also show that the experimental data can be used to estimate the individual-level repeatability of the underlying traits.

\item Our method allows experimenters to find repeatable variation in social behaviours that cannot be assessed in solitary individuals. Furthermore, experiments in animal behaviour often focus on comparisons between groups randomly sampled from a population. Increasing the behavioural variation between groups increases statistical power for testing whether a group feature is related to other properties of groups or to their phenotypic composition. Sorting according to socially relevant traits is also beneficial in artificial selection experiments, and for testing correlations with other traits. Overall, the method provides a useful tool to study how individual properties influence social behaviour.

\end{enumerate}

\vspace{0.5cm}


\section{Introduction}

Using a ruler and measuring scales, it is straightforward to assess the size and weight of an individual. And by observing or testing an individual's behaviour in isolation a number of times, we can measure some of its behavioural traits, as is now common practice in behavioural ecology~\cite{bell09,dingemanse10,wolak12}. However, social behaviour, by definition, only occurs when more than one individual interact. The extent to which individual traits can be measured in a social context depends critically on the social structure of the groups they live in~\cite{whitehead08}. This task is especially difficult in non-hierarchical groups that emerge from local interactions~\cite{sumpter10}. And while we may be able to identify differences between group-level behaviour, linking this to differences in individual phenotypes remains a significant challenge~\cite{herbert13}. This makes it difficult to properly answer important questions in biology, such as how and why social behaviour has evolved.

Some experimental methods exist to quantify social behaviour in individuals. Sociability, for example, can be assessed as a repeatable trait in fish using partitioned aquaria~\cite{brown13}. Similarly, the latency to join a group can be used~\cite{reale07}. Individuals may also be assessed within groups if these groups have a consistent network structure or division of labour: for example, social networks based on proximity or dyadic interactions can be analysed to find the most gregarious or influential individuals~\cite{whitehead08,pinter13,aplin15,spiegel16,krause16}. In other cases, repeatable differences in spatial or temporal ordering can be used to quantify individuals in terms of leader-follower or proactive-reactive traits~\cite{burns12,pettit13,pettit15,aplin14}.  In such experiments, individuals often need to be marked for identification, which can influence the expression of some behaviours~\cite{murray00}. Another potential issue is that the expression of an individual's behaviour depends on which conspecifics are available to interact with it. This is particularly important when animal groups are artificially composed by researchers.

An alternative way to investigate social behaviour at the individual level is to see how different combinations of personality types, as tested in individuals, affect the behaviour of groups~\cite{dyer09,pruitt10,brown13,hui14,planas15}. These equal-sized groups may be formed randomly, or they may be chosen to have particular phenotypic compositions. Group-level measurements may then be correlated against various descriptive statistics of the individual phenotypes comprising each group. For example, the exploratory tendency of feral guppy ({\em Poecilia reticulata}) shoals is more correlated with the lowest individual activity score and the highest sociability score than the means of either, indicating a disproportionate influence of minorities on a group's behaviour~\cite{brown13}.

One general limitation of this approach is that some behaviours are only expressed when particular social cues are available (for example, schooling or flocking responses) and hence cannot be measured in isolation, and may not be strongly correlated with personality. Another shortcoming is in the ad-hoc methods of composing groups. Randomly creating groups sampled from a larger population generally does not maximise the between-group variation in behaviour, which leads to reduced statistical power in hypothesis tests. Randomly sampled groups are statistically expected to become more similar to each other as the group size increases. The alternative of hand-picking groups with particular compositions increases power, but generally restricts the experiment to testing a few hypotheses (e.g.\ whether bold, shy, or 50-50 mixed shoals of fish forage best~\cite{dyer09}), out of several alternatives (e.g.\ shoals with one bold or one shy individual, and so on).

Here, we propose an efficient method to compose a number of animal groups that consistently behave differently to each other, such that repeatable individual differences in social behaviour can be quantified, without the need for individual labelling or assessment. Our method sorts individuals between groups, not according to their own behaviours or interactions, but on the basis of an arbitrary group-level property of interest, which we refer to here as a ``group feature''. This feature should be captured by a single continuous variable. Although there is much current interest in the evolutionary implications of group composition \cite{farine15}, this feature does not need to be directly related to fitness. For example, groups may be sorted according to cohesion, as measured by average nearest-neighbour distance. Our method is designed so that the variance in this group feature increases over time, as long as there is repeatable variation in the underlying individual behaviour: for example, different individual tendencies to aggregate. The resulting groups, ranging from high-cohesion to low-cohesion, may then be subjected to further assays in order to determine how cohesion is related to other properties of the groups.

We first detail the sorting method in Section \ref{procedure}. In Section \ref{groupmodel}, we provide a model for individual social phenotypes, and a general model for how individual contributions affect the group feature. In Section \ref{sims} we provide measures to analyse the experimental results.  In Section \ref{analytic} we define a class of group feature models for which the phenotypes become sorted by group. In Section \ref{simresults} we simulate the sorting procedure for four group feature models and analyse the data produced. We then show how to estimate the repeatability of the underlying individual traits from this data in Section \ref{estimation}. We conclude in Section \ref{discussion} by discussing how the method can provide insight into general questions in ethology, behavioural ecology and collective behaviour.

\section{Materials and methods}

\subsection{Sorting procedure}
\label{procedure}

We assume there are $N$ animals. This method is to be conducted in a laboratory setting, using animals either sampled from the wild or from a laboratory colony. Here we consider individuals that cannot be classified by other means: for example by sex or age. 

\begin{figure}[t]
\centering\includegraphics[width=80mm]{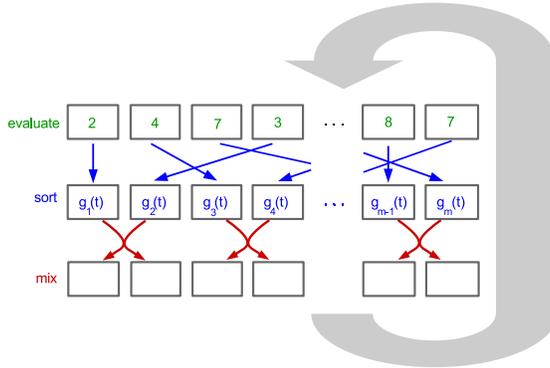}
\caption{ \label{fig1} The group sorting protocol. In each iteration, groups are assayed and then sorted according to the group feature, with the scores $g_1(t)$ to $g_m(t)$. Each pair of groups exchanges half of members, and the new groups are subject to the next round of evaluation.}
\end{figure}

The sorting is initiated by randomly dividing the $N$ individuals into $m$ groups of $n$. Without loss of generality, we assume that $m$ and $n$ are both even numbers and $N$ is chosen accordingly. Each round consists of three steps (see Fig.~\ref{fig1}). Firstly, each group is subjected to a behavioural assay, from which a score is obtained for the group feature. Such group-level scores are often obtained from the analysis of tracked videos or GPS co-ordinates \cite{herbert16}, averaging over a fixed period of time or until an objective is reached \cite{biro06,ioannou15}. Secondly, the groups are ranked from lowest to highest scoring. In the following analysis, the $j$th lowest score in each round will be referred to as $g_j(t)$. Finally, adjacent groups in this ranking exchange half of their members, chosen at random. That is, the exchange is between the 1st and 2nd ranked, between the 3rd and 4th ranked, all the way up to the pair of groups ranked $m-1$ and $m$.

The only labelling required is of the $m$ locations where the groups are housed between assays. This presents an alternative to other combinatorial optimisation algorithms, such as for group testing problems and multi-armed bandits~\cite{papadimitriou98}, which keep track of individuals as they are used in various combinations. In comparison, our algorithm is very straightforward to implement, and is also well-suited for animals that are difficult to distinguish or tag individually. The mixing stage makes it unlikely that the same exact combination of individuals appears twice, and allows the space of possible groups to be explored efficiently.

As a practical example, a selection experiment for social cohesion (i.e. small inter-individual distance) in fish would start by randomly placing, for instance, eight individuals each in 20 different tanks. These groups are then assessed by tracking the positions of individuals in each group for ten minutes, then calculating the respective average social cohesion scores, which are then ranked. In the next round, new groups are composed by swapping half of the individuals of each group with half of the groups immediately adjacent to them (above or below) on the scoreboard. Here a good catching protocol is essential, because the most daring/proactive/least-stressed individuals will often otherwise be caught first \cite{biro08}, thereby biasing group composition in the next round. To further avoid experimenter bias, the person handling the animals should be unaware of the score of the groups; also the groups should be tested in a random order. This procedure is repeated until a predefined criterion is reached i.e. a certain number of rounds, or until the ranking of the groups remains stable.

A way of assessing whether the sorting has stabilised, independently of the distribution of group scores, is by comparing the rankings between rounds. The similarity of the rankings at round $t$ to those at round $t-1$ can be measured using the correlation between the two sets of rankings. This gives a measure of group-level repeatability of behaviour. An increase in this correlation indicates that groups are changing rank by a smaller amount, and hence that the mixing between pairs has a smaller effect on the group behaviour over time. 

The mathematical analysis of the sorting procedure in this paper assumes a slight variation of the method presented above. In this version, on alternate rounds the mixing pairs are shifted by one group. That is, in these rounds mixing occurs between groups 2 and 3, groups 4 and 5, and so on. While this version is more mathematically tractable, results are not expected to differ from the experimentally simpler version presented above.

\subsection{Social phenotypes and group composition}
\label{groupmodel}
To evaluate the sorting algorithm, we assume that each individual has a phenotype, or combination of phenotypes that determines its average contribution to the group feature when assayed, and that this contribution varies randomly between evaluations. For example, an individual contribution might be ``attraction to conspecifics'', with the corresponding group feature being ``cohesion''. Mathematically, each individual is labelled $i \in \{1,...,N\}$ and has a trait value $\mu_i$. In this paper, we assume that the trait values $\mu_i$ are normally distributed with a standard deviation of one. An individual's contribution to its group's feature when evaluated at time $t$ is then a random variable $X(i,t)$  drawn from a normal distribution with mean $\mu_i$ and standard deviation $\sigma$. In this paper we will refer to $\sigma$ as within-individual variation. Due to the controlled experimental setup, the physical environment is assumed to be constant across individuals and across time. The social environment is presumed to change during sorting, however its effect on individual behaviour is not directly accessible. We assume that the remaining source of variation $\sigma$ is the same for all individuals. Although the random variation in behaviour may vary between individuals, even at the same ontogenetic stage, this assumption is parsimonious and commonly made in the statistical models that are used to estimate repeatability \cite{dingemanse10}. From these definitions, the individual repeatability can be written as \cite{bell09}
\begin{equation}
R = \frac{1}{1 + \sigma^2}
\end{equation}
Therefore $\sigma \ll 1$ indicates highly repeatable behaviour at the individual level, while $\sigma \gg 1$ indicates low repeatability.

The second important part of the model is a function that describes the group feature in terms of the contributions of its members. This may describe both the effects of the social environment on the expression of social traits \cite{webster11}, as well as how these expressions are combined in the group feature. To keep track of group membership, each group has a label $j \in \{1,...,m\}$, and we denote its set of members at time $t$ as $A^j(t)$. The group feature $g_j$ for group $j$ can therefore be written as a mathematical function, using the contributions $X(i,t)$ for every individual $i$ in the set $A^j(t)$ as the inputs. 

The sorting method does not assume any particular model for how individual contributions determine the group feature. To evaluate the method, we use some scenarios equivalent to those outlined in \cite{farine15} in terms of ``group phenotypic composition''. The simplest possibility is that the group feature is the sum or average of its parts. For example, the many wrongs hypothesis in group navigation \cite{simons04} says that the directional accuracy of a flock is determined by the average navigational ability of its members. In this case, a group feature for group $j$ can be written as
\begin{equation}
g_j(t) = M_j(t) = \frac{1}{n} \sum_{i \in A^j(t)}  X(i,t) \label{permean}
\end{equation}
A similar averaging function also fits the observed group exploration of Argentine ants~\cite{hui14}.

We may also consider a case where the median contribution drives the group feature
\begin{equation}
g_j(t) = ^\mathrm{median}_{i \in A^j(t)}  X(i,t) \label{permed}
\end{equation}
This model is expected to behave qualitatively similarly to that given by Equation \ref{permean}, but to be less sensitive to extreme phenotypes, and more sensitive to the within-individual variation of the intermediate members of groups.

Another possibility is that the group feature is largely influenced by an individual with the most extreme contribution, which plays either a leading or inhibiting role in the group. For example, the most cautious member of a group may produce social cues to make the entire group more cautious. This type of behaviour has been observed in fish, where bold individuals conform to shy conspecifics but not vice-versa~\cite{frost07}, as well as in group activity levels~\cite{brown13}. The limiting case of this model is the group depending entirely on a single ``weakest link'', in which case
\begin{equation}
g_j(t) = ^\mathrm{min}_{i \in A^j(t)}  X(i,t) \label{permin}
\end{equation}
Equivalently, the group could be limited by the ability of a single leader. Weighting in this direction has been observed in geese, where bolder individuals are more likely to make decisions \cite{kurvers10}. In the limiting case the maximum contribution is the one of interest
\begin{equation}
g_j(t) = ^\mathrm{max}_{i \in A^j(t)}  X(i,t) \label{permax} 
\end{equation}
For the analysis considered here, the above two models are equivalent.

Greater heterogeneity within a group may also increase the group feature, such as in the presence of social niches~\cite{bergmuller10}. For example, foraging species may benefit from both bold and shy individuals to promote exploration and group cohesion, respectively \cite{dyer09,michelena10}. Variation in aggression levels may also be beneficial by reducing conflict \cite{pruitt10}. A simple function approximately capturing this scenario is the standard deviation
\begin{equation}
g_j(t) = \frac{1}{n} \sqrt{\sum_{i \in A^j(t)}  \left(X(i,t) - M_j(t) \right)^2 \;} \label{perstd}
\end{equation}
where $M_j(t)$ is the mean as defined above in Equation \ref{permean}. Conversely, in other scenarios, greater homogeneity within a group increases the group feature. This may occur when interactions with similar conspecifics are stronger or more frequent \cite{croft05}. A well-known benefit of group homogeneity is the ``confusion effect'' as a defence against predation \cite{milinski84,landeau86}. If we then let the group feature $g_j$ be a measure of, for example, synchronisation, then the negative of the above formula can be used.

All of these scenarios may be considered as special cases of a general group feature function
\begin{equation} \label{groupfunc}
g_j(t) = f (X(A^j_1,t), ... , X(A^j_n,t))
\end{equation}
where $f$ is any function of the $n$ contributions.

\subsection{Simulation and measurements}
\label{sims}
We evaluate the sorting procedure based on simulations for $N=160$ individuals, divided into 10, 16 or 20 groups. Initial simulations were run with $\sigma=1$, such that the within-individual variation was equal to between-individual variation. This corresponds to a repeatability of $0.5$, which is close to the median value reported in the literature \cite{bell09}.

Simulations and subsequent analysis were run in MATLAB. Simulations were run for 50 rounds, and statistics were built up by running each simulation 5000 times. Each round produced $m$ ranked scores of the groups $g_1$ up to $g_m$. These were used to quantify the between-group variation, given simply by the standard deviation of group scores.
\begin{equation} \label{bgvar}
\sigma_g(t) = \sqrt{\sum_{j=1}^m(g_j(t) - \bar{g}(t))^2}
\end{equation}
where $\bar g(t)$ is the mean group score in round $t$. The gain is defined by the ratio of this quantity to that in the first round
\begin{equation} \label{gain}
\mathrm{gain}(t) = \frac{\sigma_g(t)}{\sigma_g(0)}
\end{equation}
This measures how much the between-group variation is increased compared to the initial random selection of groups. The group scores were transformed using the known distributions for random samples, so that initial distributions of $g_j$ were normal.

A complementary set of useful data, less dependent on the overall distribution of phenotypes, is the sequence of changes in group rank. The group ranked $j$ in round $t-1$, after swapping half of its members and being assayed again, has a new rank in round $t$ given by $k_j(t)$.  For each round of sorting, a between-rounds correlation can be calculated to quantify the stability of the rankings in that round, and thus track the sorting progress. We use the Spearman rank correlation, given by
\begin{equation} \label{corr}
C(t) = 1 - \frac{6\sum_{j=1}^m (k_j(t) - j)^2}{m(m^2-1)} \; \mathrm{for} \, t > 0
\end{equation}
which takes a maximum possible value of one if the rankings are unchanged. The expected value of $C(t)$ is zero if the rankings change randomly (i.e.~$k_j$ is entirely independent of $j$).

Using a Markovian approximation that the distribution of $C(t+1)$ is independent of $C(t)$ for all rounds, a likelihood function can be calculated for the data
\begin{equation} \label{likelihood}
L(\sigma) = \sum_{t=1}^T \mathrm{log} \, \ell_{\sigma}(t,C(t))
\end{equation}
where $\ell_{\sigma}(t,c)$ is the likelihood of the correlation in round $t$ being $c$, based on histograms obtained from performing several simulations with within-individual variation parameter $\sigma$.

\section{Results}

\subsection{Sortability for monotonically increasing group feature}
\label{analytic}

Before we look at simulations of our sorting algorithm, we prove that the algorithm will be effective for a wide class of group feature functions (equation \ref{groupfunc}). Specifically, many biologically realistic group features are such that replacing a member of a group with one of a higher trait value will not decrease the expected group feature. Mathematically, this means the group feature function is weakly monotonic, or non-decreasing on the individual contributions. These functions includes the mean (equation \ref{permean}) and other weighted averages, as well as the minimum (equation \ref{permin}), median (equation \ref{permed}) and maximum (equation \ref{permax}), and other functions based on quantiles or thresholds. This class does not include cases where the group feature depends on homogeneity, as in equation \ref{perstd}.

To make our analysis independent of the distribution of $\mu$, we use the global ranks of the individual trait values. Let the global ranks of the individuals in group $j$ in round $t$ be $A^j_1(t)$ through $A^j_n(t)$ in ascending order, and let
\begin{equation}
S_j(t) = \sum_{i=1}^n  A^j_i(t)
\end{equation}
be the sum of ranks in group $j$ in round $t$. By our definition above, the group scores $g_j(t)$ are increasing with $j$. As the group scores diverge over time, we expect the rank-sums $S_j$ to be similarly diverging for monotonic functions $f$. In the Supporting Information we prove that this is the case for weakly monotonic group feature functions.

The theorem implies that over time the sorting algorithm increases the within-group homogeneity and between-group diversity in terms of the individual phenotypes. If the monotonicity condition does not hold, then it is not guaranteed that within-group variation of phenotypes will decrease, even if between-group variation increases.

\subsection{Simulation results}
\label{simresults}
\begin{figure}[t]
\centering\includegraphics[width=140mm]{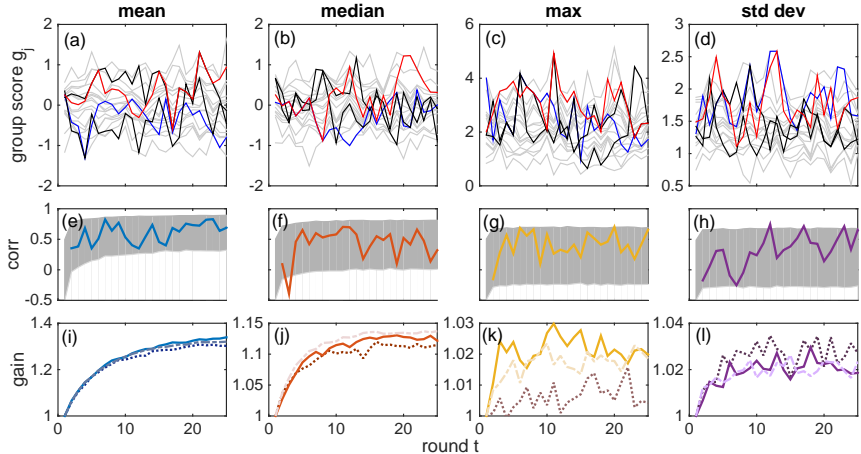}
\caption{\label{fig2}Typical simulations of the sorting procedure for four group feature models.
(a-d) Ranked group scores over 25 rounds of sorting using (a) the mean, (b) the median, (c) the maximum and (d) the standard deviation of the individual contributions as the group feature, using $m=16$ groups of $n=10$ individuals. To show how the individual phenotypes move between groups over time, we highlight the group that contains the individual with the highest (red), lowest (blue) and the two median (black) phenotypes out of all $N$ individuals. (e-h) Correlation coefficients between the set of group ranks before mixing and the set after mixing, corresponding to the simulations above. Grey areas represent the 95\% confidence intervals. (i-l) Expected between-group variance, normalised to the initial variance, for $m=10$ groups of $n=16$ individuals (dotted lines), $m=16$ groups of $n=10$ individuals (full lines) and $m=20$ groups of $n=8$  individuals (dashed lines). In all simulations, the within-individual variation is $\sigma=1$, corresponding to a repeatability of $R=0.5$}
\end{figure}

The changes in group feature scores over time are shown in Fig.~\ref{fig2}(a-d) for typical simulations of four group feature models, based on the mean (equation \ref{permean}), the median (equation \ref{permed}), the max (equation \ref{permax}) and the standard deviation (equation \ref{perstd}) of the phenotypes. The individuals with the overall most extreme and median hidden phenotypes $\mu_i$ are tracked as they move between groups, indicating how they are sorted over time. Panels (e-h) show the corresponding between-round correlations in group ranks given by equation \ref{corr}, and panels (i-l) show the average gain in between-group variation given by equation \ref{gain}. Both of these measures, even for the non-monotonic, `standard deviation' model, increase up to an equilibrium value. The gain in between-group variation, as well as the maximum between-round correlation, are largest for the `mean' model. 

Both the global statistics and the sorting of the hidden phenotypes depend on the group feature model. When the group feature is proportional to the mean of contributions (Figure~\ref{fig2}(a,e,i)), the initial between-group standard deviation is $\sigma_g(t=0) \approx \sqrt{(1 + \sigma^2) /n}$. Over time, individuals will tend to become grouped according to the underlying ranks of their phenotypes. For fully sorted groups, the between-group variation should approach the between-individual variance of 1. For all three group sizes the standard deviation of the group feature increases by about 33\% after 25 rounds. For the `median' model (Figure~\ref{fig2}(b,f,j)), the dynamics are similar, but less efficient in increasing the between-group gain and the between-round correlation.

For the `maximum' group feature model, shown in Figure \ref{fig2}(c,g,k), the highest scoring group always contains the individual with the highest contribution in that round, and over time the individuals with highest trait values $\mu_i$ become clustered together in this group.  Under this scenario, the process is highly sensitive to within-individual variation, which slows down the overall sorting procedure. While the individuals with low $\mu_i$ are not well correlated with the groups they are associated with, those individuals with high $\mu_i$ (i.e.~those that dominate group behaviour) are sorted together in higher-ranked groups. While not shown in Figure \ref{fig2}, the opposite case of minimum-based or "weakest link" limited groups (equation \ref{permin}) is identical to this, only with all group features and individual trait values changing sign. In this case, the individuals with the lowest trait values become sorted into the lower-ranked groups. 

In the `mean', `median', `min' and `max' models, the individuals become approximately sorted according to their phenotypes, as predicted by the result in Section \ref{analytic}. When the group feature increases with heterogeneity in the group , i.e.~the group feature is proportional to the standard deviation, this is no longer the case (Figure \ref{fig2} (d,h,l)). While between-group variation still increases, this is due to both tails of the trait distribution becoming mixed in the same groups.

Keeping the total number of individuals constant and using a different group size has little effect on the between-group variation over 25 rounds (Figure \ref{fig2} (i-l)). In all cases, the group size affects the gain by less than 5\%. For all models and group sizes, the majority of the increase in between-group variation occurs within the first 15 rounds.

\begin{figure}[t]
\centering\includegraphics[width=140mm]{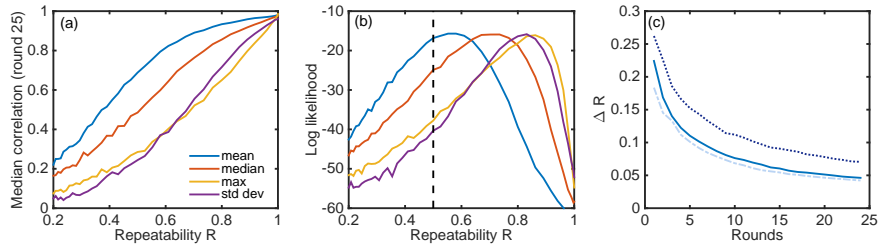}
\caption{ \label{maxlik} Estimating the individual repeatability. (a) The median correlation in group rankings after 25 rounds, out of 1000 simulations, as a function of the individual repeatability R, for the four group feature models. Simulations are for 16 groups of 10. (b) Typical log-likelihood functions for the repeatability generated by a single experiment with actual value $R=0.5$ (dashed line) and a `mean' group feature model. (c) Standard deviation of the conservative estimate of the repeatability (`mean' model) as a function of the number of rounds for $R=0.5$. Dotted line: 10 groups of 16; solid line: 16 groups of 10; dashed line: 20 groups of 8.}
\end{figure}

\subsection{Estimation of individual-level repeatability}
\label{estimation}

The stability of the group rankings between rounds is dependent on the number of rounds of sorting that have passed, as well as the model for the group feature, as well as the individual repeatability (controlled by the within-individual variation $\sigma$). Figure \ref{fig2}(e-h) shows how the between-rounds correlation depends on time and on the group feature for a constant repeatability. Figure \ref{maxlik} (a) shows how the typical correlation close to equilibrium (25 rounds) increases with repeatability for each model. This is because in all cases, more predictable individuals lead to more predictable groups.

Using the between-rounds correlation $C(t)$ as a statistic for each round, a maximum likelihood method can be used to estimate the repeatability.  For every model and for every combination of parameters, 1000 simulations were run and histograms were compiled to represent the likelihood of a correlation $C(t)$ for each round $t$ (i.e.~$\ell_{\sigma}(t,c)$ from Equation \ref{likelihood}). Figure \ref{maxlik} (b) shows the result of a single simulated experiment which is fit against these models. In the experiment, the individual repeatability was $R=0.5$ and the group function was the mean. Fitting against the mean model, the estimate obtained by maximum-likelihood is close to the correct value. The alternative `maximum', `standard deviation' and `median' models produce a similar likelihood, but estimate a much higher repeatability. 

The `mean' model, out of all considered, gives the most conservative estimate for the repeatability. Running multiple experiments, the uncertainty of this estimation can be quantified by the standard deviation of the estimates, shown by Figure \ref{maxlik} (c). The uncertainty decreases with the number of rounds as more data is collected, with greater uncertainty for larger groups. In all cases, the uncertainty of the conservative estimate is reduced below $0.1$ within 15 rounds. This uncertainty has only a slight dependency on the underlying repeatability and model, as shown in Supporting Information.

\section{Discussion}
\label{discussion}

We have proposed a simple and general method for increasing variation in behaviour at the group level, and for studying the individual basis of socially relevant behaviours. Whatever is the dependence between group features and individual contributions, the phenotypic composition of the top and bottom groups naturally approach the optimum compositions for extreme group features. This avoids the need to exhaustively test many different compositions, which for larger groups is only feasible with agent-based simulations~\cite{michelena10,aplin15}. For a wide class of group feature functions, the individuals become increasingly sorted according to their trait values. This result is robust to the distribution of trait values and, as the sorting is driven by these expected phenotypes, we expect it to also be robust to between-individual variation in plasticity. For the `mean' model, where all individuals contribute equally to the group feature, this sorting results in a substantial increase in the between-group variation.

Our approach is especially useful for animals that aggregate in self-organised groups such as fish shoals, since it does not require individual labelling or tracking behaviour of individuals within groups. It is in such self-organised groups that differences between individuals are the most obscured. Only a small number of sorting rounds (10-20) is required to achieve groups that are close to maximally stable for all models considered here. For even relatively small groups ($n=8$) the total number of assays required is comparable to the number required to separately assay each individual twice, as is commonly done when estimating repeatability~\cite{bell09}. However, attention to welfare issues may be required, as individuals are subject to more handling by this method. We now highlight five areas of application where we believe our method can be particularly useful.

Increased between-group variation enables more statistical power when investigating the link between group behaviour and group phenotypic composition~\cite{dyer09,pruitt10,kurvers10,brown13,hui14,planas15}. This may be done after sorting, also without labelling, by assaying the individuals from each group. For instance, the members of the top group may be split into $n$ arenas for assaying, followed by the next group, and so on. Regression or correlations can then reveal how certain summary statistics for the $n$ assays (for example, the mean or standard deviation) are dependent on the group ranking.  In principle, information about the ideal group composition can also be gleaned from the sorting data, using a more detailed model selection than is shown here. However, we expect that it may be difficult to distinguish the `maximum' and `standard deviation' models, for example, which evolve in a similar way during sorting. In addition, some measures, such as the gain in between-group variation, may not be reliable test statistics if habituation to the assay is to be expected.

Our method can be used to confirm and quantify consistent individual-level differences in socially relevant behaviour. This has the potential to add several behaviours, which are only expressed in groups, to the behaviours in which repeatable variation is known to exist~\cite{bell09}. If the repeatability is very low, the group rankings are expected to be random every round. An increase in the between-rounds correlation of the group rankings over time may be used as evidence of repeatable social behaviour. Without knowledge of the correct group feature model, the `mean' model provides a conservative estimate of this repeatability. If the group feature model can be inferred, for example using the method above, a more exact estimate of the repeatability can be made.

Investigating the effects of group composition can also be used to test mechanistic models of collective behaviour. The basic parameters of these models, such as the size and weighting of alignment and attraction zones, are often approximated via detailed analysis of micro-level interactions over time~\cite{sumpter12,cavagna10,mann14,herbert11}. These models often further predict certain relationships between group phenotypic composition and macroscopic group-level behaviour~\cite{couzin02,conradt05}, but these individual differences are hard to identify in large groups. By comparing the `rules of interaction' in groups with different features, it can be seen what rules are covarying with the group feature and with each other, which may provide insight into the perceptual basis of collective motion~\cite{strandburg13,herbert16}. 

Identification of other traits that are correlated with social behaviour has also been difficult due to the earlier identified problems with detecting individual qualities in a collective setting. Our method now allows for high throughput analysis of individual aspects of social behaviour and sorting of individuals for later analysis of, for example, morphological traits \cite{partridge93,kotrschal13} and life-history traits \cite{hunt06}. Correlations with other behavioural traits may also be used to investigate behavioural syndromes~\cite{sih12}. As we have shown, the individual phenotypes can be sorted as long as the group feature is a monotonic function of all member contributions.

Our method also provides a basis for the experimental design of artificial selection experiments on generic social behaviours.  This opens up a new experimental route to studying the evolvability and genetic basis of social behaviour~\cite{falconer96}. Artificial selection can be performed by breeding individuals from a top and/or bottom quantile of the $m$ sorted groups. If the actual dependence of the group feature on individual contributions is monotonic (as in the mean, median, minimum or maximum cases), this should result in directional selection. In the case where the group feature increases with the heterogeneity of the group (as in equation \ref{perstd}), the top and bottom trait values will be mixed together in the groups with highest feature levels. The lowest-scoring groups in this case can be used for artificial stabilising selection.

Our analysis can be extended to the sorting of multiple traits with different influences on group-level behaviour. If the group feature is expected to be a linear combination of functions on independent traits (e.g.~ if $g_j$ increases with the mean of phenotypes of trait X and decreases with the variance of phenotypes of trait Y in the group) then the sorting is expected to operate on both traits independently, with relative efficiency depending on the relative repeatability of traits X and Y. However, if the group feature depends on an interaction between phenotypes of X and Y across individuals (e.g.~ if $g_j$ the product of the two means of the phenotypes of X and Y) then we expect results to differ from what is presented here, since multiple distinct combinations of phenotypes may result in the same group feature. This is a fruitful topic for further study.

Various other extensions can be made to the models and analysis presented in this paper. A more detailed model selection can be done using the full set of rank transitions, rather than just the correlation of rankings between rounds. Moreover, since our method is quite general, the analysis we have presented can be extended to include other group feature models. A time-dependence may be added to the modelled group behaviour in order to account for habituation or changes in the physical environment. A time-dependence could also be used to model social learning \cite{griffiths03,croft04}, although the randomised exchange of group members in each round is expected to mitigate this effect.

To conclude, we have generated a general and efficient experimental method for studying the individual basis of group level behaviour. Not only does this algorithm produce data useful for inference, but it also lends itself to many otherwise elusive experiments in collective behaviour, making it possible to bridge the empirical gap between individual properties and collective outcomes.

\noindent {\bf Acknowledgements.} AS would like to thank Oliver Johnson and Melinda Babits for helpful suggestions. The authors would also like to thank multiple referees of previous versions of this manuscript for substantial suggestions for improvement. This work was funded by The Knut and Alice Wallenberg Foundations grant 102 $2013.0072$.

\noindent {\bf Data Accessibility.} MATLAB code used for simulations and for maximum likelihood inference is available in Supporting Information.

\nolinenumbers

\bibliographystyle{unsrt}

\newpage

\section*{Supplementary Information}
\section*{A: Proof of sortability result}

We consider $N$ individuals being sorted in $m$ groups of $n$. The $N$ hidden individual trait values are constant in time, and are written in increasing order as $\{ X(1),..., X(N) \} \in {\mathbb R}$.

Let the global ranks in group $j$ in round $t$ be $A^j_1(t)$ through $A^j_n(t)$ in ascending order

The groups in each round are sorted by their group score
\begin{equation}
g_j(t) = f(X(A^j_1(t)), .. ,X(A^j_n(t)))
\end{equation}
such that $g_j(t)<g_{j+1}(t)$ for all $j<m$ and $t$.

In odd rounds, mixing occurs between groups $2k-1$ and $2k$ for all $1 \leq k \leq m/2$, while in even rounds, mixing occurs between groups $2k$ and $2k+1$ for all $1 \leq k < m/2$. This is assumed to be done by combining into a group of size $2n$ then randomly partitioning back into groups of size $n$.

Let 
\begin{equation}
S_j(t) = \sum_{i=1}^n  A^j_i(t)
\end{equation}
be the sum of ranks in group $j$ in round $t$. We expect these to be increasing with $j$ along with $g_j$ if the groups are well sorted. Let
\begin{equation}
Q_j(t) = P(S_j(t) > S_{j+1}(t)) \, \forall \, j<m
\end{equation}
be the probability that groups $j$ and $j+1$ are in the wrong order in terms of the rank sum.

{\bf Proposition 1:}
When there is mixing of groups $j$ and $k$ and there is no rearranging between pairs of groups, the sum $S_j + S_k$ is conserved.

{\bf Proposition 2:}
Furthermore, if $f$ is monotonic we know $\mathrm{E}[S_j (t+1)]$ and $\mathrm{E}[S_k (t+1)]$ are both monotonically increasing functions of $S_j(t) + S_k(t)$.

{\bf Proposition 3:}
When $f$ is monotonic
\begin{equation}
 \forall \, j,k,t : g_j(t) < g_k(t) \; \exists r : A^j_r(t) < A^k_r(t) \;
\end{equation}
If $A^j(t)$ and $A^{j+1}(t)$ are random partitions of the same set at time t, this implies the upper bound 
\begin{equation}
Q_j(t) \leq \frac{1-1/n}{2}
\end{equation}

This therefore applies to all groups at $t=0$ i.e. 
\begin{equation}
E[S_j(0)]<E[S_{j+1}(0)] \; \forall \, j < m
\end{equation}
since the starting groups are randomly sampled from the population. This observation also applies to each pair of groups upon mixing.

{\bf Theorem 1:}
{\em If f is nontrivial and non-decreasing on all arguments, then}
\begin{equation}
Q_j(t+2) \leq \lambda Q_j(t) \; \forall \, j \in \{2,...,m-2\} , t
\end{equation}
{\em for some $\lambda < 1$ and therefore on average the groups become more homogenous over time}

{\bf Proof:}
Consider $m=4$ groups. At round $t$, groups 2 and 3 are mixed. We know from proposition 3 that $\mathrm{E}[S_2(t-1)] < \mathrm{E}[S_3(t-1)]$ and $\mathrm{E}[S_2(t)] < \mathrm{E}[S_3(t)]$. If the pair is anomalous at time $t-1$, that is, $S_2(t-1) > S_3(t-1)$ (which has probability $Q_2(t-1)$) invoking proposition 1 leaves us with
\begin{equation}
\mathrm{E}[S_2(t) - S_2(t-1)] < 0
\end{equation}
and
\begin{equation}
\mathrm{E}[S_3(t) - S_3(t-1)] > 0
\end{equation}

In round $t+1$, groups 1 and 2 are mixed, and groups 3 and 4 are mixed. In this anomalous case, the sum $S_1 + S_2$ is expected to decrease, and the sum $S_3 + S_4$ is expected to increase, since $S_1$ and $S_4$ were unchanged. Using proposition 2, we know that $S_2(t+1)$ is expected to decrease, and $S_3(t+1)$ is expected to increase. Therefore
\begin{equation}
P[S_2(t+1) > S_3(t+1) | S_2(t-1) > S_3(t-1)] < P[S_2(t-1) > S_3(t-1)]
\end{equation}
In the other case at $t-1$, there is no expected change to $S_2$ and $S_3$.
\begin{equation}
P[S_2(t+1) > S_3(t+1) | S_2(t-1) < S_3(t-1)] = P[S_2(t-1) < S_3(t-1)]
\end{equation}
By the law of total probability, when $Q_2$ is nonzero, it must decrease by a nonzero amount every two rounds. This proof is easily expanded to $m>4$ groups.

\newpage

\section*{B: Repeatability and sorting progress}

\begin{figure}[!h]
\centering\includegraphics[width=125mm]{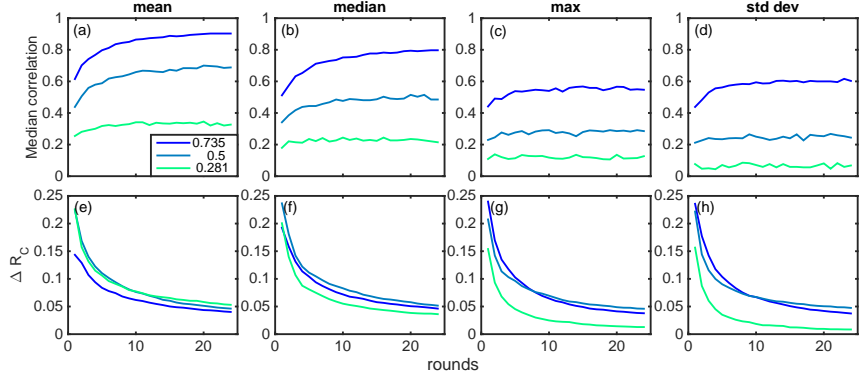}
\caption{ \label{repfig} Expected sorting progress for different values of repeatability. (a-d) Median between-rounds correlation and (e-h) standard deviation in the conservative repeatability estimate vs round number, for the four group feature models, for 16 groups of 10. For each set of parameters, 1000 simulations were run.}
\end{figure}

To investigate the effect of changing repeatability on the speed of sorting, we ran simulations for $\sigma=0.6$, $\sigma=1$ and $\sigma=1.6$, corresponding to repeatability values of $R=0.735$, $R=0.5$ and $R=0.281$ respectively. For lower values of $R$, the equilibrium between-rounds correlation is lower, but is reached in a shorter time compared to larger $R$. As more data is gathered with each round, the uncertainty in the conservative estimate of $R$ decreases by a similar amount regardless of the underlying chosen model and repeatability. For the `maximum' and `standard deviation' models with low repeatability, the uncertainty in the estimate is decreased. This may be an artifact due to the estimated values being close to zero and hence close to the boundary of possible repeatability values.

\end{document}